\documentclass[twocolumn,showpacs,preprintnumbers,amsmath,amssymb]{revtex4}

\usepackage{graphicx}% Include figure files
\usepackage{dcolumn}% Align table columns on decimal point
\usepackage{bm}% bold math

\begin{document}

\preprint{APS/123-QED}

\title{The effect of electron beam pitch angle and density gradient on solar type III radio bursts}% Force line breaks with \\

\author{R. Pechhacker}
 %\email{r.pechhacker@qmul.ac.uk}%Lines break automatically or can be forced with \\
\author{D. Tsiklauri}%
 %\email{d.tsiklauri@qmul.ac.uk}
\affiliation{%
School of Physics and Astronomy, Queen Mary University of London, E1 4NS, United Kingdom
}%

\date{\today}% It is always \today, today,

\begin{abstract}
1.5D Particle-In-Cell simulations of a hot, low density electron beam injected into magnetized, maxwellian plasma were used to further explore the alternative non-gyrotropic beam driven electromagnetic emission mechanism, first studied in Ref.\cite{2011PhPl...18e2903T}. Variation of beam injection angle and background density gradient showed that the emission process is caused by the perpendicular component of the beam injection current, whereas the parallel component only produces Langmuir waves, which play no role in the generation of EM waves in our mechanism. Particular emphasis was put on the case, where the beam is injected perpendicularly to the background magnetic field, as this turned off any electrostatic wave generation along the field and left a purely electromagnetic signal in the perpendicular components. The simulations establish the following key findings: i) Initially waves at a few $\omega_{ce}/\gamma$ are excited, mode converted and emitted at $\approx \omega_{pe}$ ii) The emission intensity along the beam axis is proportional to the respective component of the kinetic energy of the beam; iii) The frequency of the escaping EM emission is independent of the injection angle; iv) A stronger background density gradient causes earlier emission; v) The beam electron distribution function in phase space shows harmonic oscillation in the perpendicular components at the relativistic gyrofrequency; vi) The requirement for cyclotron maser emission, $\frac{\partial f}{\partial v_{\perp}} > 0$, is fulfilled; vii) The degree of linear polarization of the emission is strongly dependent on the beam injection angle; viii) The generated electromagnetic emission is left-hand elliptically polarized as the pitch angle tends to $90^{\circ}$; ix) The generated electromagnetic energy is of the order of $0.1\%$ of the initial beam kinetic energy.
\end{abstract}

\maketitle

\section{\label{sec:intro}Introduction}
Solar type III radio bursts are believed to be a result of relativistic electrons of $\approx 10-100$keV, that propagate through solar plasma. The electrons might be accelerated in impulsive solar flares as they are observed to occur in groups of 10 and more \cite{1971AuJPh..24..201M}. Potential sources for acceleration include magnetic reconnection and dispersive Alfven waves \cite{2011PhPl...18i2903T,2012PhPl...19h2903T}, which are both likely to be found in flare regions.\\
Observationally, the electromagnetic emission spectrum of type III bursts show a characteristic drop of frequency over time in the 'dynamical spectrum' (frequency versus time intensity plot). The frequency of the observed electromagnetic signal typically drops a few orders of magnitude within minutes. The emission occurs at two separate frequencies, the fundamental plasma frequency and its second harmonic. It is observed that type III bursts are accompanied by electrostatic Langmuir waves as well as evidence for electron distribution functions that show a bump in the forward direction (parallel to the background magnetic field) in phase space \cite{1970SoPh...12..266L}. Further, there is evidence for back propagating Langmuir waves as well as ion acoustic waves.\\
The common theory suggests that fast electrons, that propagate outward on an open magnetic field line through the Sun's atmosphere, trigger the 'bump-on-tail' instability, which is responsible for the generation of the observed Langmuir waves. The Langmuir waves then, in turn, generate electromagnetic emission at the local plasma frequency and its second harmonic. As the beam of fast electrons propagates from denser into less dense regions, the frequency of the generated emission drops, because the plasma frequency is a function of the square root of the density. There are various theories on how the electromagnetic emission is generated. Those theories include non-linear wave-wave interactions (interactions of Langmuir waves with backward propagating Langmuir waves or ion-acoustic waves)\cite{1958SvA.....2..653G,1985JGR....90.6637C}, linear mode conversion on density gradients (partial reflection and partial conversion of a Langmuir z-mode wave)\cite{2008PhPl...15j2110K}, and the antenna mechanism (trapping of Langmuir waves in density cavities to drive currents at the second harmonic)\cite{2012ApJ...755...45M,2010JGRA..11501101M}.\\
Recent works \cite{2011PhPl...18e2903T,2005ApJ...622L.157S} studied type III burst scenarios in particle-in-cell (PIC) type simulations. Ref.\cite{2005ApJ...622L.157S} investigated a magnetic field setup near a magnetic reconnection region in a 2.5D (2D in space, 3D in velocities, fields),  Ref.\cite{2011PhPl...18e2903T} considered a generic magnetic field line connecting Sun to earth in a 1.5D. Simulations in Ref.\cite{2011PhPl...18e2903T} were carried out with EPOCH, a fully relativistic, electromagnetic PIC code. In the present study of solar type III radio bursts, EPOCH was also used to simulate different beam injection angles into magnetised plasmas, as well as the effect of different background density gradients on the generated emission. The simulations were set up to inject a mildly relativistic, hot, low density beam into a spatially 1D plasma, that featured a constant background magnetic field as well as a parabolic density profile, both parallel to the spatial axis. The default setup and analysis tools were tested by successfully reproducing the results published in Ref.\cite{2011PhPl...18e2903T}. Ref.\cite{2011PhPl...18e2903T} established that generation of electromagnetic emission via a non-gyrotropic beam injection is possible. The primary goal of this study is the effect of different beam pitch angles and density gradients on the electromagnetic wave generation, as well as proving that EM emission can be generated without invoking Langmuir waves or the classical plasma emission mechanism, which involves Langmuir and ion-sound waves satisfying resonant beat conditions to mode convert into escaping transverse EM radiation.\\ 
In order to investigate the effect of different beam injection angles, components of the beam injection momenta were varied, while the total beam momentum (and hence the total kinetic energy of the beam) was kept constant in all cases. Results showed that a parallel beam momentum resulted in generation of an electrostatic Langmuir wave, visible in the parallel electric field component, travelling at the speed of the beam (essentially Langmuir turbulence following the beam). There was no excitation of Langmuir waves found, when the beam is injected perpendicularly to the background magnetic field. In the perpendicular components of electric and magnetic fields, waves, that travel approximately with the speed of light, were found. This electromagnetic signal was also found for a perpendicular beam injection. Hence, the simulated generation of electromagnetic emission seems independent of the generation of electrostatic Langmuir waves.\\
It should be noted, that, in 1.5D geometry, the classic plasma emission mechanism is switched off, because it involves interaction of Langmuir, ion-acoustic and electromagnetic waves, that, when treated mathematically, involves a factor of $|{\bf k_L} \times {\bf k_T}|^2$, with ${\bf k_L}$ being wave vector of the Langmuir wave and ${\bf k_T}$ for the transverse electromagnetic wave. The angle between the two wave vectors for the 1.5D case is equal to zero, setting also the interaction probability to zero.\\
In the case of perpendicular beam injection, the electron distribution function will not show a bump in the forward direction, therefore, the instability, that is supposed to be responsible for the Langmuir wave generation, is never created. This result suggested an alternative emission mechanism.\\
Closer analysis of the distribution function in (parallel and perpendicular) momentum space showed, that, for non-parallel beam injection, a positive gradient was to be found in the perpendicular direction. Such shapes of distribution functions are referred to as cyclotron maser or loss-cone distributions. In special cases, if distribution functions are shaped appropriately, they are also called shell or horseshoe distributions, as studied in Ref.\cite{1999JGR...10410317P,2000ApJ...538..456E} in the context of auroral kilometric radiation (AKR). It is known that these distributions allow for wave generation in most prominently x- and o-mode, but also z-mode \cite{1986ApJ...307..808W}. Which mode is growing strongest is prescribed by the ratio $\omega_{ce}/\omega_{pe}$, i.e. how the magnetic field compares to plasma density, where $\omega_{ce}$ is the electron gyrofrequency and $\omega_{pe}$ is the plasma frequency.\\
This paper is structured as follows: In Sec.\ref{sec:simset} we will discuss in detail the simulation setup. We will vary the beam injection angle in Sec.\ref{sec:varangle} and demonstrate the impact of different background density gradients in Sec.\ref{sec:vardens}. In Sec.\ref{sec:disfunc}, we will study the behaviour of the distribution function, while, in Sec.\ref{sec:pol}, we will analyse the polarization of emitted waves as well as relate the generated field energy to the initial beam kinetic energy. Finally, we will summarize key points in \ref{sec:conc}.

\section{\label{sec:simset}Simulation Setup}
All simulations, presented in this paper, use EPOCH, a fully electromagnetic, relativistic particle-in-cell code that was developed by the Engineering and Physical Sciences Research Council (EPSRC)-funded collaborative computational plasma physics (CCPP) consortium of UK researchers.\\
We are considering a 1.5D maxwellian plasma, allowing spatial variation in $x$ only, while electromagnetic fields and particle momenta have all three components. The background magnetic field along this line is kept constant $B=B_x=0.0003$T$=3$G, setting the electron gyrofrequency to $\omega_{ce}=5.28 \times 10^7$Hz rad everywhere. The background temperature is $T=3 \times 10^5$K and isotropic. The background plasma density at $x=0$ is $n_0=10^{14}$m$^{-3}$, giving $\omega_{pe}=5.64 \times 10^8$Hz rad. This sets $\frac{\omega_{ce}}{\omega_{pe}}=0.0935 \ll 1$. Further, the electron Debye length is $\lambda_{De}=3.78 \times 10^{-3}$m.\\
The simulation setup is such that we simulate a single magnetic field line connecting Sun and earth, while the grid size is $\lambda_{De}/2$. It is predicted that in the limit $r \gg R_{\bigodot}$ the plasma density $n_e(r) \propto r^{-2}$ (\cite{1999A&A...348..614M,1998SoPh..181..429R}), therefore,  we make use of

\begin{equation}\label{eq:normaln}
  n_{e,i}(x) = n_0 \left[ \left( \frac{x-x_{max}/2}{x_{max}/2 + n_+} \right)^2 + n_- \right]%
\end{equation}

with $x_{max}$ being the total system length and

\begin{eqnarray}
  n_+ = \frac{x_{max}}{2} \frac{1-\sqrt{1-n_-}}{\sqrt{1-n_-}}, & n_- = 10^{-8} %
\end{eqnarray}

resembling a parabolic density drop off, such that $n_{e,i}(0) = n_{e,i}(x_{max}) = n_0$ and $n_{e,i}(x_{max}/2) = n_- n_0$. The unrealistic growing density in the region $x_{max}/2 \leq x \leq x_{max}$ is employed in order to use periodic boundary conditions. When analysing results we, therefore, draw conclusions from the region $0 \leq x \leq x_{max}/2$ only.\\
Further, in section \ref{sec:vardens}, we consider modified plasma density profiles to study the effects of different density gradients. The following density profiles are used to study a \textit{weak} and \textit{strong} gradient

\begin{eqnarray}
  n_{w}(x) = n_0 \left[ \left( \frac{x-x_{max}/2}{x_{max}/2 + n_+} \right)^2 + n_- \right], &  n_- = 0.5\\%
  n_{s}(x) = n_0 \left[ \left( \frac{x-x_{max}/2}{x_{max}/2 + n_+} \right)^8 + n_- \right], &  n_- = 10^{-8}.%
\end{eqnarray}

\begin{figure}
\includegraphics[scale=0.49]{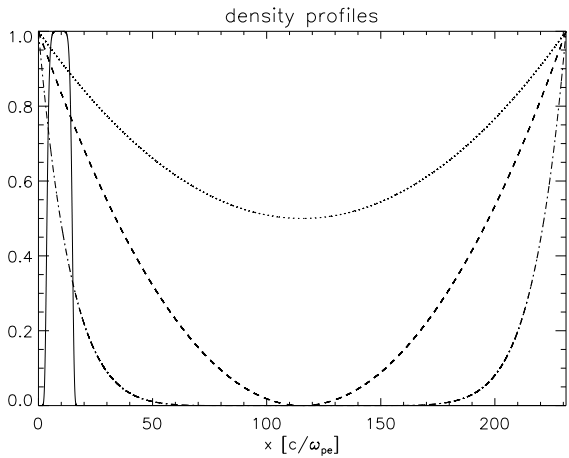}%
\caption{\label{fig:densprofs} The electron beam (solid), weak (dotted), moderate (dashed), strong (dotted-dashed) density gradients at $t=0$. Densities are normalized to their maxima.}
\end{figure}

A beam of accelerated electrons is injected at simulation time $t=0$. It carries a total momentum of $p_b=m_e \gamma \frac{c}{2}$ with $\gamma \approx 1.155$. The beam temperature is $T_b=6 \times 10^6$K. The peak beam density is $n_{b0}=10^{11}$m$^{-3}$, while its spatial profile is defined as

\begin{equation}
  n_b(x) = n_{b0} e^{-[(x-x_{max}/25)/(x_{max}/40)]^8}%
\end{equation}

setting the beam density maximum at $x_{max}/25$. In order to demonstrate the situation at $t=0$ in terms of densities as a function of space, we plot the beam and background number density profiles in Fig.\ref{fig:densprofs}. Note that the beam is injected only once, at $t=0$, i.e. the beam electrons are not replenished.

As a result of the above defined quantities, at $x=0$, the plasma beta is $\beta=0.0115$. The mass ratio used is $m_i/m_e=1836$.\\
Each simulation is run on 32 Dual Quad-core Xeon processors with 4 Tb of RAM. The simulation makes use of 65000 grid points and 10000 particles per cell in the case of the background electrons and ions. Despite the low beam electron density ($n_{e,i}(0)/n_b = 10^3$), the simulation achieves a reasonable number of beam electrons per cell, since the beam is focussed on only roughly 10 simulation cells, resulting in 100 beam electrons per cell. Herein lies, however, a shortcoming of the simulation setup as realistic background-to-beam electron ratios of $\approx 10^5 - 10^7$ cannot be achieved, because they would require considerably larger computational resources.

\section{\label{sec:varangle}Varying Beam Injection Angle}
In this section we analyse the effect of a variation of the beam injection angle (with respect to the background magnetic field) on the EM emission. We change the injection pitch angles via variation of a perpendicular component in the beam momentum, $p_{by} = p_b \sin \theta$, while keeping the total momentum $p_b$ constant in all cases. Therefore, an increase of the beam pitch angle $\theta$ is equivalent to an increase of perpendicular beam momentum $p_{by}$, respectively a decrease in parallel beam momentum $p_{bx} = p_b \cos \theta$, hence, a decrease in propagation speed. For the case $\theta=90^{\circ}$, the parallel component of the beam momentum vanishes and the beam is trapped at its point of injection. The background plasma density profile in this section is always represented by Eq.(\ref{eq:normaln}) which corresponds to the dashed line in Fig.\ref{fig:densprofs}. We study cases of $\theta=15^{\circ},45^{\circ},60^{\circ},90^{\circ}$. 
In order to analyse the simulation data, we make use of time-distance plots and spatial wavelet transforms of the electromagnetic field components.
Time-distance plots offer the possibility of studying the development of a physical quantity with respect to space and time. We make use of this technique to plot components of electric and magnetic fields as well as the electron density. We observe waves in both electric and magnetic components. The slope that is being produced by the wave front in this plot can then shed light on the propagation speed of that wave. We choose units for spatial and temporal axis such that a slope of $\approx 1$ corresponds to propagation with the speed of light. Hence, finding waves that travel with $\approx c$ suggests that the emission is electromagnetic.
Dynamical spectra show a characteristic feature of solar type III bursts. These plots show the development of the emission frequency over time. In the case of solar type III bursts, the dynamical spectrum shows a frequency drop over time. This drop is due to the decreasing background plasma density, that a beam of accelerated electrons encounters, while travelling away from the Sun. The drop in density correlates to a drop of the local plasma frequency, since $\omega_{pe} \propto \sqrt{n_{e}}$. As the beam triggers emission at the local plasma frequency (and its second harmonic), while it travels from denser into less dense plasma, it generates emission of dropping frequency that is characteristic for solar type III bursts.
We investigate the frequency of the generated emission by performing a wavelet analysis of a perpendicular component of the electric field; we choose $E_y$. We take a snapshot of $E_y$ at a given time and perform a spatial wavelet analysis. The wavelet software was provided by C. Torrence and G. Compo\citep{22}.

\begin{figure*}
\includegraphics{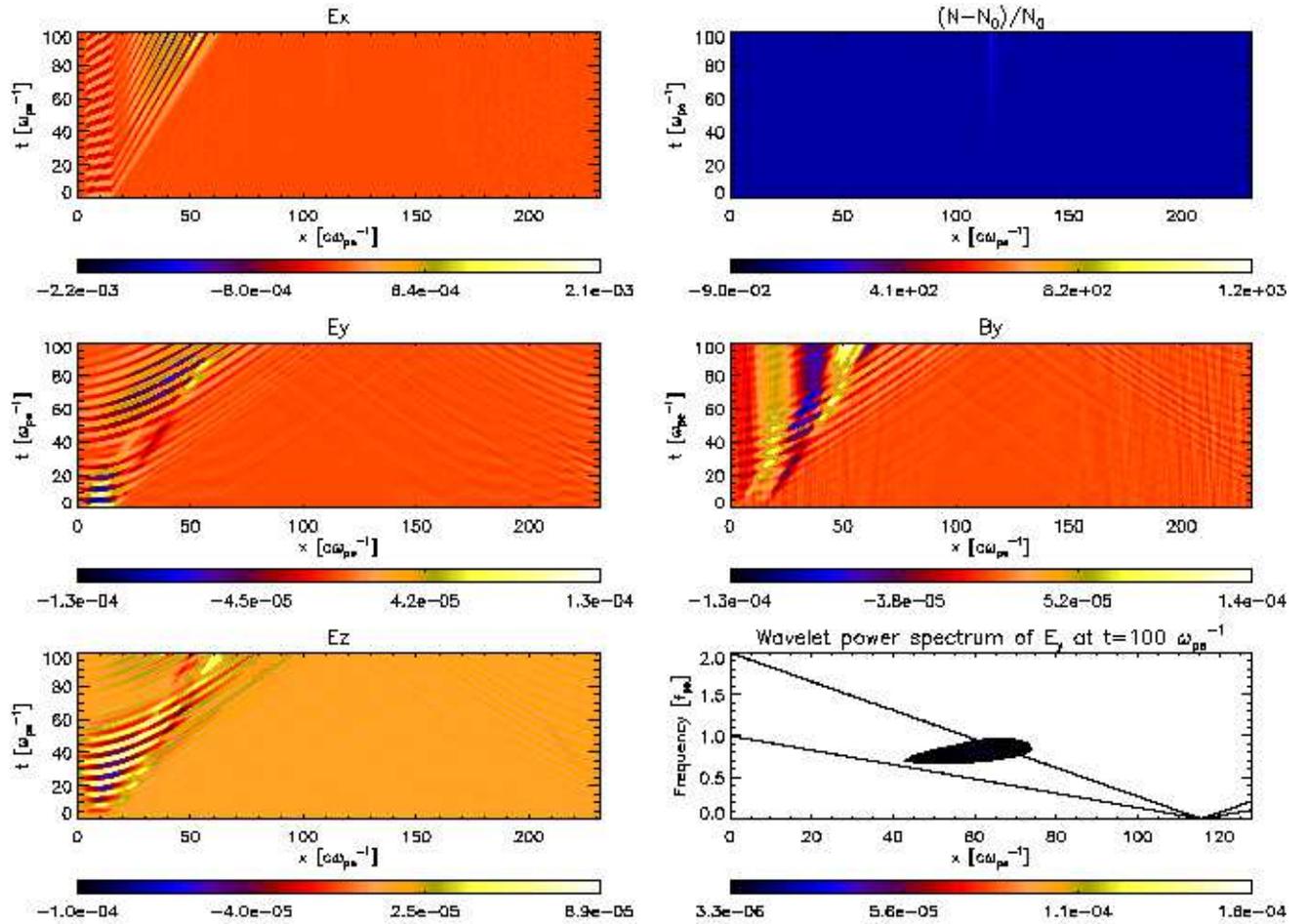}
\caption{\label{fig:tdds15} $\theta=15^{\circ}$. The moderate density profile, given by Eq.(\ref{eq:normaln}), was used. Left column: time-distance-plots of electric field components. Right: time-distance-plot of changes in density, magnetic field $y$-component, and the wavelet transform of $E_y$. Note that the background for the wavelet transform was set to white colour and does not refer to maximum amount of emission on the sides of the plot. Further, the black lines track the local plasma frequency and its second harmonic. }
\end{figure*}

\begin{figure*}
\includegraphics{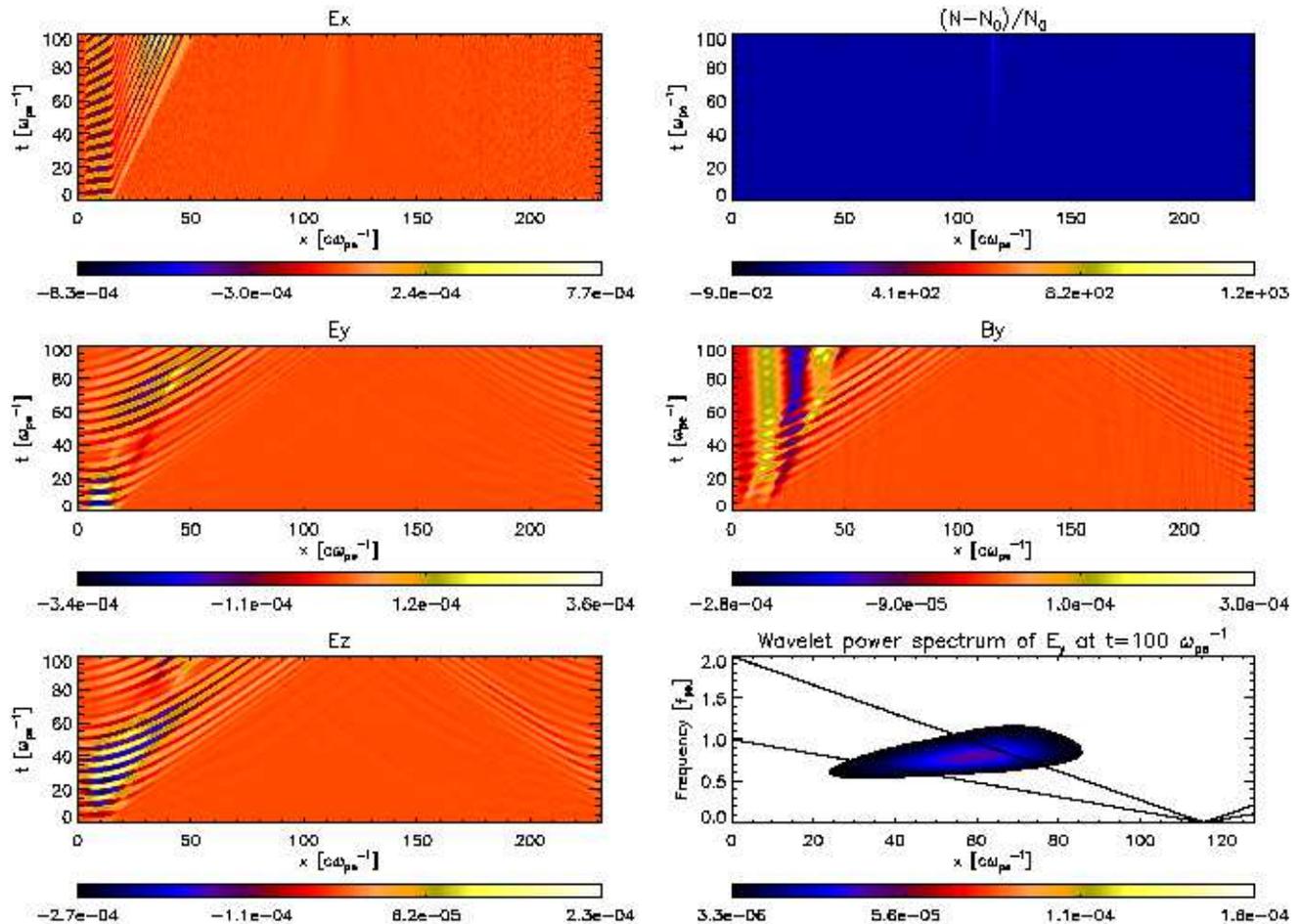}
\caption{\label{fig:tdds45} As Fig.\ref{fig:tdds15}, but $\theta=45^{\circ}$.}
\end{figure*}

\begin{figure*}
\includegraphics{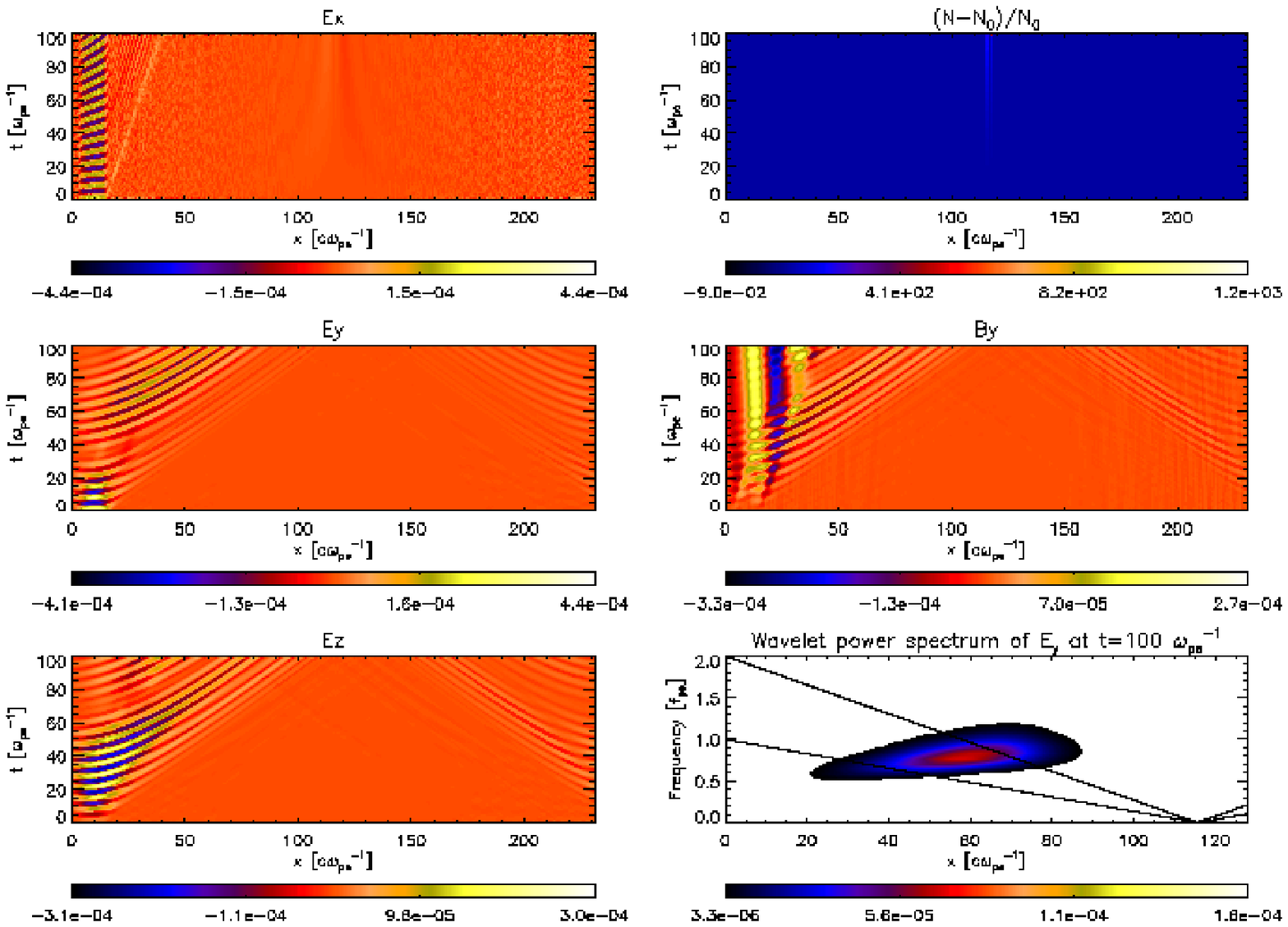}
\caption{\label{fig:tdds60} As Fig.\ref{fig:tdds15}, but $\theta=60^{\circ}$.}
\end{figure*}

\begin{figure*}
\includegraphics{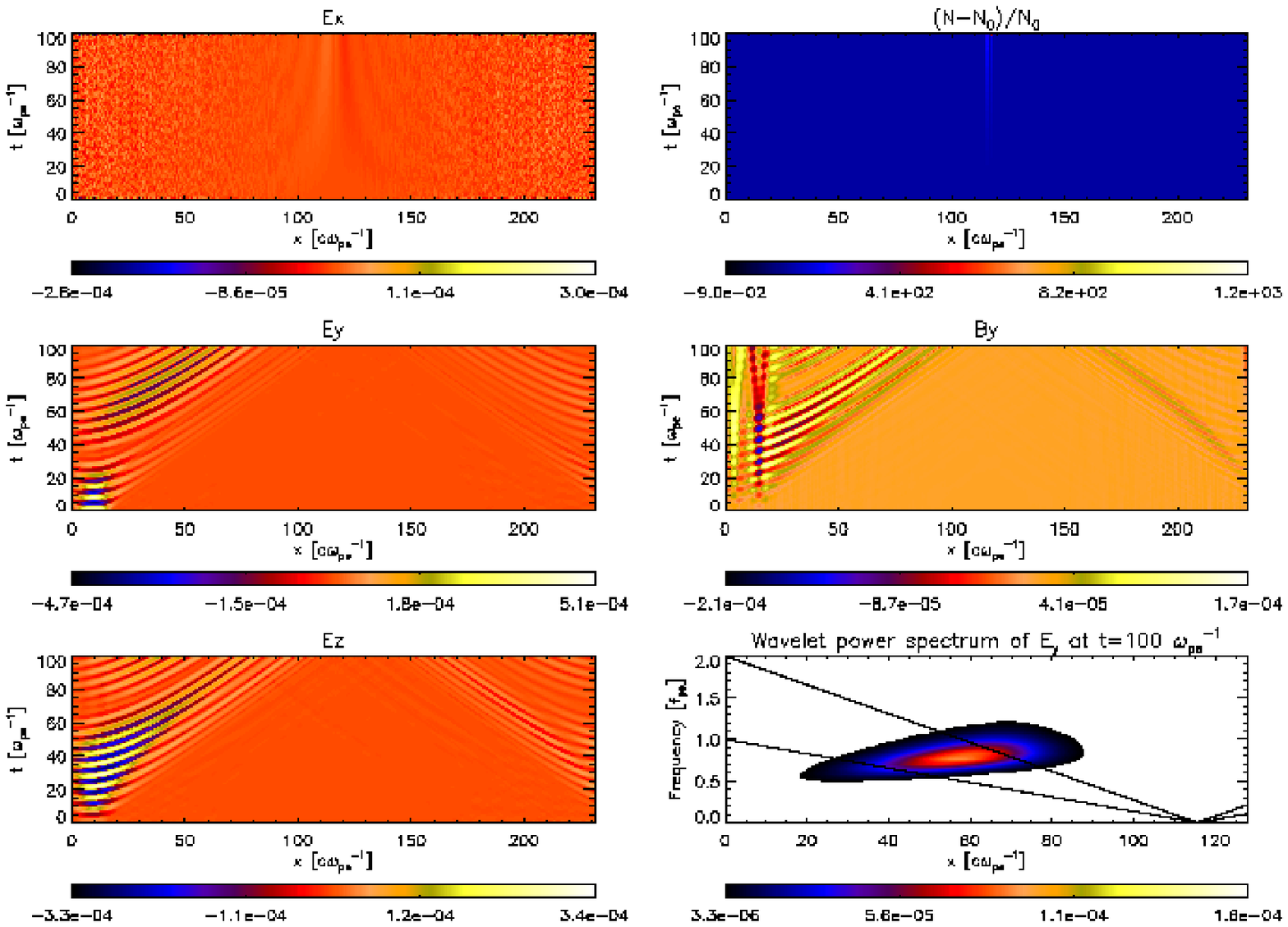}
\caption{\label{fig:tdds90} As Fig.\ref{fig:tdds15}, but $\theta=90^{\circ}$.}
\end{figure*}

\begin{figure}
\includegraphics[scale=0.49]{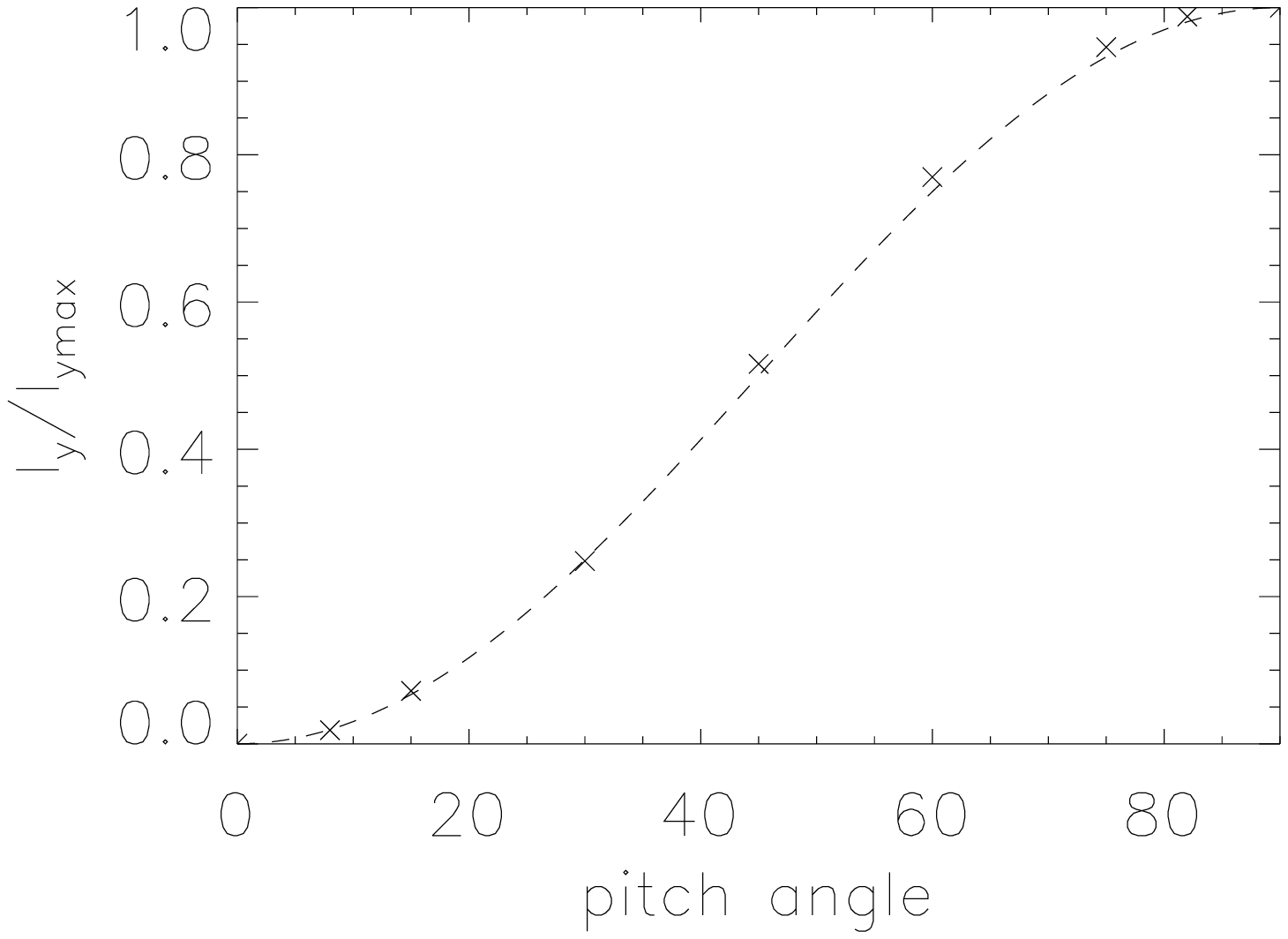}
\caption{\label{fig:IoverTheta} The maximum emission intensity $I_{max}(\theta)$ normalized to $I_{max}(90^{\circ})$ (crosses) and $\sin^2(\theta)$ (dashed) as a function of pitch angle $\theta$ in degrees. }
\end{figure}

Figs.\ref{fig:tdds15},\ref{fig:tdds45},\ref{fig:tdds60},\ref{fig:tdds90} show time-distance plots of components of the electric field in the left column. The right column contains plots of changes in number density, the magnetic $y$-component and the wavelet transform of $E_y$ taken at $t=100 \omega^{-1}_{pe}$. Electric and magnetic field strengths are normalized to units of $\omega_{pe}cm_e/e$ and $\omega_{pe}m_e/e$, respectively, while distance and time are measured in $c/\omega_{pe}$ and $\omega^{-1}_{pe}$, accordingly, where $\omega_{pe}$ is the plasma frequency at $x=0$. This is a good value of reference, as the background density (and therefore the plasma frequency) at this point are equal in all simulations, regardless of which density profile for the background (see Fig.\ref{fig:densprofs}) is used.\\
Comparing the time-distance-plots for electromagnetic fields, we find wave generation in the perpendicular components for all injection angles. The wave front shows a slope of $\approx 1$, is therefore travelling with the speed of light and can be interpreted as electromagnetic emission. The wave propagation speed in the perpendicular components does not depend on the beam injection angle and is therefore independent of the beam propagation speed. Similar analysis of the parallel component $E_x$ shows rather different behaviour. In the $\theta=90^{\circ}$ case (Fig.\ref{fig:tdds90}), $E_x$ does not show any wave features and no parallel component to the emission is generated. In the other cases (Fig.\ref{fig:tdds15},\ref{fig:tdds45},\ref{fig:tdds60}), we can see generation of a standing wave as well as a travelling electrostatic wave, which shows different propagation speeds for each of the cases, and can be associated with the beam propagation speed. This signal also has components in the perpendicular plane, most notable in the magnetic $B_y$ component, but also in (rather weakly) $E_y$ and $E_z$, where it interferes with contributions of the electromagnetic emission. The width of the signal at $t=0$ corresponds to the width of the beam at the start of the simulation.\\
In all cases, the density profile remains unchanged (see top right panels on Figs.\ref{fig:tdds15}-\ref{fig:tdds90}). This is important as it excludes any possibility of developing density cavities in the plasma, which could trap Langmuir waves and contribute to the emission in form of the antenna mechanism \cite{2010JGRA..11501101M}. Thus, in our simulations, EM emission is generated by a mechanism other than the antenna mechanism or plasma emission (the latter requires at least 2D in space).\\
Comparing wavelet transforms for various pitch angles, we clearly see that emission intensity is proportional to the perpendicular component of the beam momentum. Other than that, there is no evidence of change in the wavelet transform and emission characteristics as peak frequencies, emission times (respectively location) and wave packet shapes are all preserved. Analysis of the peak emission intensity as a function of pitch angle is shown in Fig.\ref{fig:IoverTheta}.

The figure shows that the peak emission intensity (normalised to its maximum value at $\theta=90^\circ$) can be almost perfectly approximated by $\sin^2(\theta)$, and therefore $p^2_{by}$. This result suggests a direct correlation of the beam kinetic energy, $E^b_{kin} \propto p^2_b$ and the generated emission intensity.

\section{\label{sec:vardens}Time Development and Varying the Background Density Profile}
\begin{figure*}
\includegraphics{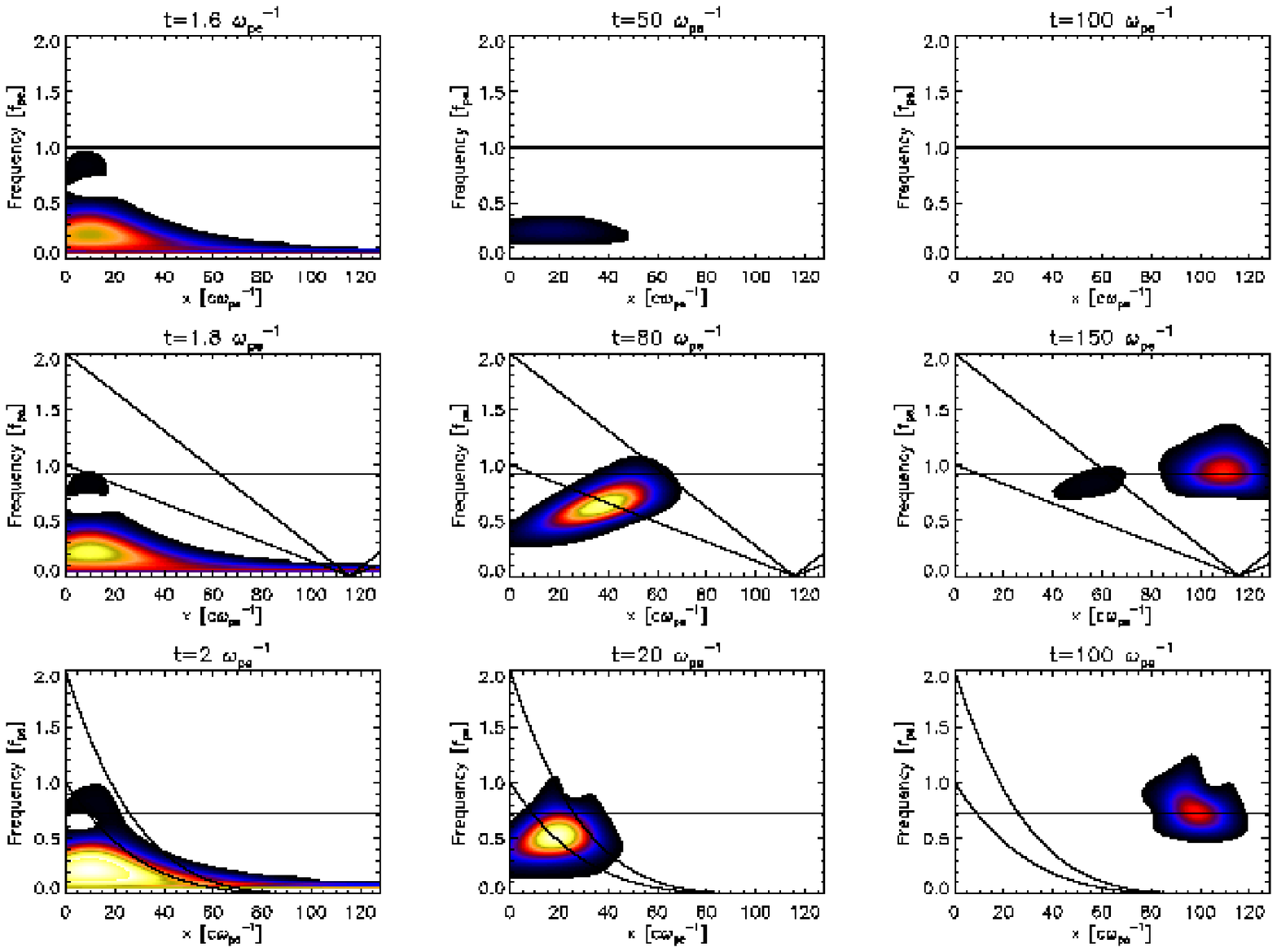}
\caption{\label{fig:dscomp} Wavelet transforms for different density gradients at different times. Top row: density is constant. Middle row: moderate gradient according to dashed line in Fig.\ref{fig:densprofs}. Bottom row: strong gradient according to dotted-dashed line in Fig.\ref{fig:densprofs}. Black curves correspond to local plasma frequencies and the second harmonic, the horizontal line is the plasma frequency corresponding to the point of the beam density maximum at the time of the beam injection, $f_{inj}$. }
\end{figure*}

We now focus on the $\theta = 90^{\circ}$ case, as it shows the maximum generated emission in $E_y$. In this section we investigate the role of the background density profile. We refer the reader to Fig.\ref{fig:densprofs} for the different density profiles. Additionally, we investigate a case without any density drop, where $n_c(x)=n_0$.\\
The time development of the wavelet transform allows insight into the emission generation process, wave propagation, as well as frequency drift. We generated movies to demonstrate this (movie 1, movie 2, movie 3, and movie 4 in Ref.\cite{mov:ref}). Again we focus on the wavelet transform of $E_y$. We include curves that mark the local plasma frequency and its second harmonic, as well as  $\omega_{z}=-\frac{1}{2}\omega_{ce}+\frac{1}{2}\sqrt{\omega_{ce}^2+4 \omega_{pe}^2}$, which is the cut-off for the z-mode. From the movies, it is evident that as soon as the beam is injected, we observe a pulsating signal of generated emission. The frequency of this emission is much smaller than $\omega_{pe}$ and, therefore, does not allow instant propagation (note that $\omega_{ce}/\omega_{pe} \ll 1$). Instead, pulsation eventually dies down only before an apparently stable wave packet is formed. The stable packet then drifts to higher frequencies. This suggests mode coupling of the cyclotron emission on the gradient to (possibly) a z-mode. When the peak frequency of the wave packet is comparable to the local plasma frequency (which for our parameters is not significantly different from $\omega_{z}$), the packet starts propagating. Interestingly, the increase in frequency does not stop at this point but only when the wave packet travels into a region where the second harmonic of the local frequency, $2f_{local}$, matches the fundamental of the plasma frequency at the injection point $f_{inj}$. (Note that the beam is not injected at $x=0$, but has a finite width and $f_{inj}$ is the plasma frequency at the beam density function's maximum at $x_{max}/25$.) From this point on the wave packet moves on with constant frequency.\\
In Fig.\ref{fig:dscomp}, we show snapshots of wavelet transforms at different simulation times. We choose to look at an early snapshot of the pulsation phase, a second snapshot when the wave packet couples to frequencies $\approx f_{pe}$ and, ultimately, the final snapshot in the simulation. Note that for the mid row case, the total simulation time has been extended to $t=150 \omega^{-1}_{pe}$. In order to not get any interference due to periodic boundary conditions, the total system size was increased by a factor of $1.5$ in this run.

We see that for constant density, there is no emission, as the wave packet never mode couples to frequencies that would allow propagation. 
At this point, we sum up the key findings:
The beam injection evokes pulsating emission far below the local plasma frequency $f_{pe}$. Eventually, this cyclotron emission mode couples to the z-mode, which has a frequency comparable to the local plasma frequency, and that allows for escape of the emission. The wave packet frequency increases up to the point where it reaches the plasma frequency of the original beam injection point, $f_{inj}$. This happens at the point where the second harmonic of the local plasma frequency, $2f_{local}$, is of the order of $f_{inj}$.\\
These observations enable us to draw the following conclusions with regards to the role of the density profile in our simulations:
In case of the constant density profile, there is no escaping emission, as the generated emission cannot couple to a propagating wave mode.
A steeper gradient in the density profile will result in earlier emission, because the mode coupling process couples to a frequency that is able to escape sooner than it does for weaker gradients.\\
We would like to note that Fig.\ref{fig:dscomp} as well as all movies of wavelet transforms in Ref.\cite{mov:ref} suggest that the pulsation maximum is not at the y-axis value corresponding to $f_{ce}$. This is due to the fact that in order to be able to interpret the y-axis as frequency, one needs to be able to relate space and time via $x=ct$. The latter is only applicable to escaping electromagnetic radiation. In the left bottom corner of the wavelet transforms, i.e. the beam injection region, this relation does not hold, as we are dealing with a trapped mode far below $f_{pe}$. Therefore, whilst the wave packet resides in this region, the y-axis actually corresponds to the wave number $k$ rather than the wave frequency. Numerical runs of wider (respectively narrower) beam width, not shown here, corroborate that the y-value of the spatial wavelet transform in regions well below $f_{pe}$ scales as $k$. It is only after $f \approx f_{pe}$ and $x=ct$ applies, that the y-axis can be interpretted as wave frequency. Further, the fact that the initial pulsation is excited at the relativistic cyclotron frequency can be evidenced from figures equivalent to Figs.\ref{fig:tdds15},\ref{fig:tdds45},\ref{fig:tdds60},\ref{fig:tdds90} but for longer runs, which are not shown here. $E_y$ pannels of the figures shown here already suggest that there is a time interval of $\approx 60 \omega_{pe}^{-1}$ between the bottom left wave structure and the one following. Thus, the frequency of the oscillation is $1/\Delta t = 2 \pi f_{pe}/60 \approx 0.1f_{pe} \approx f_{ce}$.

\section{\label{sec:disfunc}Distribution Function Dynamics}
\subsection{\label{sec:dfkart}Cartesian Coordinates}
At any given timestep, in EPOCH, distribution functions for every species are functions of space and momentum, i.e. $f(x,{\bf p})$. They satisfy the equation

\begin{equation}
  \int f_{\sigma}(x,{\bf p})d^3p dx = N_{\sigma}%
\end{equation}

with $N_{\sigma}$ being the total number of particles of a species $\sigma$ (electrons, ions, beam electrons).
The distribution functions of both the background plasma and the beam electrons in momentum space are initially maxwellian. This allows us to separate the spatial from the momentum part, by $f(x,{\bf p})=\tilde{f}(x) \tilde{f}({\bf p})$, at $t=0$. The momentum part can, then, be described by

\begin{equation}
  \tilde{f}_{e,i 0}({\bf p}) = \tilde{n}_{e,i} e^{-(p_x^2+p_y^2+p_z^2)/(2 m_{e,i} k_BT)}%
\end{equation}

for the background and

\begin{equation}\label{eq:dfbeam_p}
  \tilde{f}_{b 0}({\bf p}) = \tilde{n}_{b} e^{-[(p_x-p_{bx})^2+(p_y-p_{by})^2+p_z^2]/(2 m_{e} k_B T_b)}%
\end{equation}

for the beam respectively, where $k_B$ is Boltzmann's constant and $\tilde{n}_{e,i,b}$ are normalization constants corresponding to a species (electrons, ions, beam electrons).\\

EPOCH allows for distribution functions to be computed with respect to spatial and momentum coordinate components, as well as all desired combinations. It is therefore possible to generate arrays for distribution functions in all momentum components, $f(p_x,p_y,p_z)$, but also specific components, for example $f(p_x)$,$f(p_x)$... Here, the spatial dependency has already been integrated out by EPOCH. Following the time development of the distribution function for various angles, we refer to movie 5 and movie 6 in Ref.\cite{mov:ref}. In the movies we show the development of the distribution function in three panels, top to bottom: $f(p_x)$, $f(p_y)$, $f(p_z)$.

\begin{figure}
\includegraphics[scale=0.49]{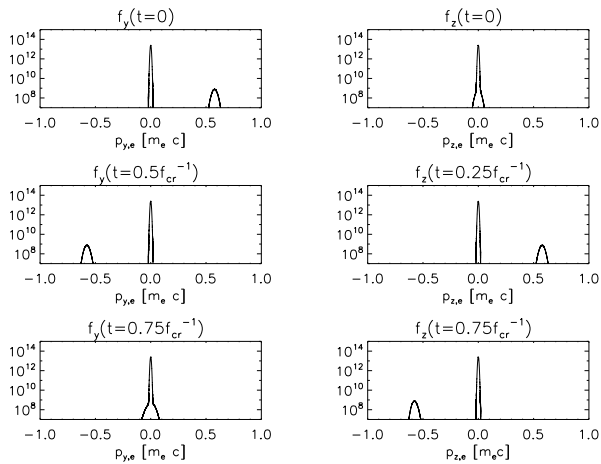}
\caption{\label{fig:dfcomp} Distribution function for the background and beam electrons in momentum space at different times (from top to bottom) for $\theta=90^{\circ}$; $f(p_y)$ (left column) and $f(p_z)$ (right column) }
\end{figure}

\begin{figure}
\includegraphics[scale=0.49]{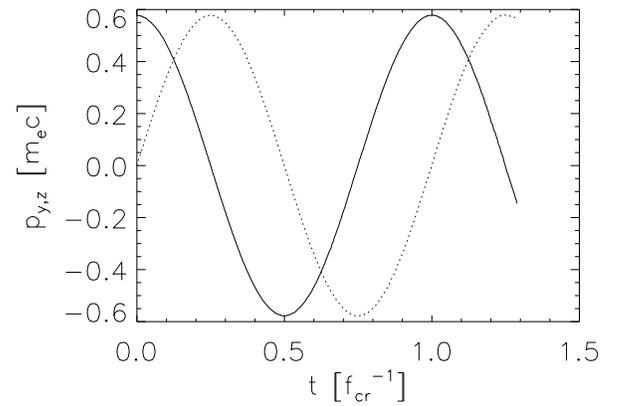}
\caption{\label{fig:dfosc} Position of beam maximum in momentum space as function of time for $\theta=90^{\circ}$; $p_y$ (solid) and $p_z$ (dotted) in units of $[m_ec]$. }
\end{figure}

Fig.\ref{fig:dfcomp} shows snapshots of $f(p_y)$ and $f(p_z)$ at different times. Each panel, generally, shows two bumps. The larger one refers to the background electrons, while the smaller one corresponds to the beam. We can clearly see that the background distribution function as well as the beam distribution in \textit{parallel} momentum space ({\bf B} along $x$) remain almost constant throughout our simulations. In \textit{perpendicular} momentum space, however, we observe oscillation of the beam distribution, as well as a broadening and widening of the beam bump. We focus on the perpendicular panels and follow the position of the beam maximum in time, which is shown in Fig.\ref{fig:dfosc}.

Fig.\ref{fig:dfosc} is representative for all cases studied. The maxima for this run are at $\frac{\gamma}{2} \approx 0.5775 m_ec$. (Note that for different injection angles, the maxima are shifted to different values on the ordinate, because different injection angles result in different values of $p_y$, but the functional characteristics will stay the same.) It shows that the maximum of the beam distribution function oscillates in the perpendicular components. The frequency of the oscillation is the relativistic cyclotron frequency, $\omega_{cr}=\omega_{ce} / \gamma$. It is trivial to deduce that the beam is gyrating around the magnetic field. The evident gyration along with the observation, that the initially generated pulsating emission is of the order of the cyclotron frequency.

\subsection{\label{sec:dfpp}Parallel and Perpendicular Coordinates}

While the cartesian form offers a valuable look into dynamics of the beam, investigation of instabilities often require a transformation of the distribution function to a momentum space with coordinates parallel and perpendicular to the background magnetic field, i.e. $f(p_{\parallel},p_{\perp})$. While in the more conventional plasma emission process, Langmuir waves are being generated via the 'bump-on-tail'-instability, referring to a bump in the distribution function in \textit{parallel} momentum space, $\frac{\partial f}{\partial p_{\parallel}}>0$, the requirement for the triggering of cyclotron maser (or loss-cone) emission is a positive slope of the distribution function with respect to the \textit{perpendicular} direction, $\frac{\partial f}{\partial p_{\perp}}>0$. EPOCH does not allow for direct output of $f(p_{\parallel},p_{\perp})$, so a transformation algorithm was developed. The transformation relations are straight forward

\begin{eqnarray}
  p_{\parallel} = p_x \nonumber \\
  p_{\perp} = \sqrt{p_y^2 + p_z^2} %
\end{eqnarray}

\begin{figure}
\includegraphics[scale=0.49]{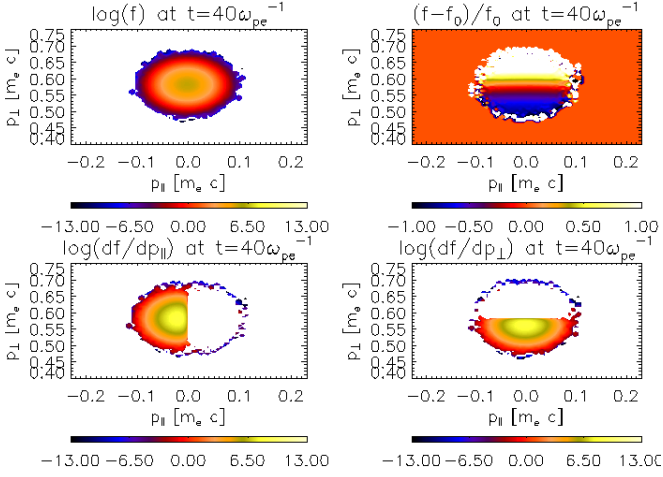}
\caption{\label{fig:dfpp} Top left: $f(p_{\parallel},p_{\perp},t)$. top right: $\frac{f(t)-f(0)}{f(0)+\epsilon}$. bottom left: $\frac{\partial f(t)}{\partial p_{\parallel}}$. bottom right: $\frac{\partial f(t)}{\partial p_{\perp}}$ at $t=40\omega_{pe}^{-1}$. Gradients are shown on logarithmic scales, which cannot show negative values, therefore, they appear white as the background. }
\end{figure}

The data volume accumulated by the simulation is prescribed by the resolution of the distribution function. A resolution of 300 was chosen for an interval of $-0.91 m_ec < p_{x,y,z} < 0.91 m_ec$, to retain reasonable resolution while keeping data volume comparatively small. The algorithm was tested by successfully reproducing Fig.1 in Ref.\cite{1986ApJ...307..808W}.\\
Movie 7 in Ref.\cite{mov:ref} shows the time development of the distribution function for the $\theta=90^{\circ}$ case. The top left panel shows a plot of the distribution function on a logarithmic scale at a given time step, while the top right panel plots the relative change with respect to the initial distribution, $\frac{f(t)-f(0)}{f(0)+\epsilon}$ (where $\epsilon$ is a small number to avoid division by zero), in order to visualise dynamics. In the lower row, the gradients $\frac{\partial f}{\partial p_{\parallel}}$ (left) and $\frac{\partial f}{\partial p_{\perp}}$ (right) are shown. We can clearly see that the initial distribution fulfils the criterion $\frac{\partial f}{\partial p_{\perp}}>0$. Further, the movie shows interesting dynamics in the shape of $f(p_{\parallel},p_{\perp})$. The left panel shows no drastic changes throughout the simulation time, while the relative change on the right panel suggest shifting to higher and lower perpendicular momenta. First, the distribution increases in momenta that lie below the initial maximum, $\approx 0.5775 m_ec$, then the movement is reversed. Until around $t \approx 30\omega_{pe}$ we can see frequent change in the direction of shifting. Comparing this to movie 3 in Ref.\cite{mov:ref}, we see that the time scales correspond to the pulsating phase in the wavelet transform, suggesting an energy exchange between beam and EM fields, as found in Ref.\cite{2012ApJ...752...60Y}. Eventually the distribution shifts to momenta above the initial maximum and
forms almost a straight line at $\approx 0.5775 m_ec$, separating regions of increased distribution from decreased ones, at $t \approx 40\omega_{pe}$. Comparison with movie 3 in Ref.\cite{mov:ref} shows that this is about the time, when we can see a stable (no longer pulsating) wave packet being formed. At $t \approx (70-90)\omega_{pe}^{-1}$, the distribution function shifts again to momenta below $\approx 0.5775 m_ec$. The wavelet transform shows that this is the time, when the wave packet passes over the local plasma frequency limit. Fig.\ref{fig:dfpp} is a representative snapshot at $t=40\omega_{pe}^{-1}$. The gradients in the bottom row are presented on a logarithmic scale, therefore, negative gradients cannot be shown and appear in the same white background colour.

\section{\label{sec:pol}Generated Electromagnetic Field Energy and Polarization}
We calculate the electromagnetic field energy by use of the Poynting theorem,

\begin{eqnarray}
  w(t) = \int [ \frac{1}{2} \epsilon_0 {\bf E}(x,t)^2 + \frac{1}{2 \mu_0} {\bf B}(x,t)^2 ] dx%
\end{eqnarray}

We take care to exclude contributions from the background magnetic field $B_0$. We relate the field energy to the initial kinetic energy of the beam, $E^{beam}_{kin}(0)$, which we calculate analytically using the expression for the mean relativistic kinetic energy

\begin{eqnarray}\label{eq:ekinrel}
  <E^{beam}_{kin}(0)> =  \int n_b(x) m_e c^2[\gamma(<{\bf p}_0>) - 1] dx%
\end{eqnarray}

where $\gamma(<{\bf p}_0>)$ is the Lorentz factor in terms of initial bulk momentum

\begin{eqnarray}
  \gamma(<{\bf p}_0>) = \sqrt{1 + (\frac{<{\bf p}_0>}{m_e c})^2} %
\end{eqnarray}

with the initial bulk momentum $<{\bf p}_0>$, given by

\begin{eqnarray}
  <{\bf p}_0> = \frac{\int {\bf p} f_0(x,{\bf p}) d^3p}{n(x)} \nonumber \\
= \frac{\tilde{f}_0(x) \int {\bf p} \tilde{f}_0({\bf p}) d^3p}{n(x)}
\end{eqnarray}

Using the relation

\begin{equation}
  \int f(x,{\bf p})d^3p = n(x)%
\end{equation}

along with the normalization condition

\begin{equation}
  \int \tilde{f}({\bf p})d^3p = 1%
\end{equation}

we see that the spatial part of the distribution function becomes the particle density distribution

\begin{equation}
  \tilde{f}(x) = n(x)%
\end{equation}

\begin{figure}
\includegraphics[scale=0.5]{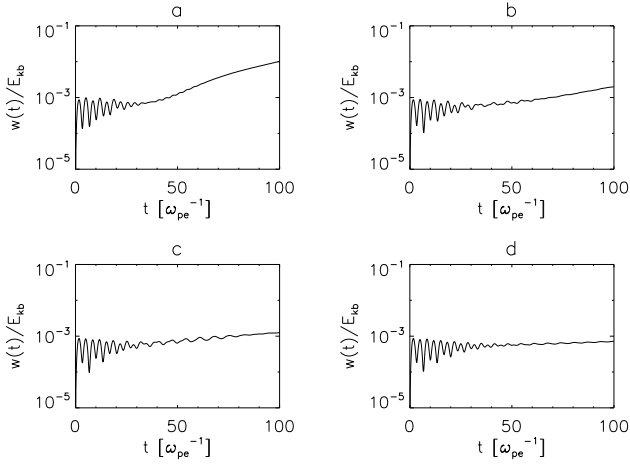}
\caption{\label{fig:EmofEk} Total field energy generated, $w$, normalized to initial kinetic energy of the beam, $E^{beam}_{kin}(0)$ according to Eq.\ref{eq:ekinrel}, as a function of time for a) $\theta=15^{\circ}$ b) $\theta=45^{\circ}$ c) $\theta=60^{\circ}$ d) $\theta=90^{\circ}$ }
\end{figure}

Hence, the bulk momentum can be written as

\begin{eqnarray}
<{\bf p}_0> = \int {\bf p} \tilde{f}_0({\bf p}) d^3p %
\end{eqnarray}

where $\tilde{f}_0({\bf p})$ is the initial beam distribution function and can be deduced from Eq.(\ref{eq:dfbeam_p}). For our simulation parameters, the initial kinetic energy of the beam is $E^{beam}_{kin}(0)=7.89272 \times 10^{-3}$J. We use this result to see how much of the beam energy gets converted into magnetic energy, by plotting this in Fig.\ref{fig:EmofEk}.

\begin{figure}
\includegraphics[scale=0.5]{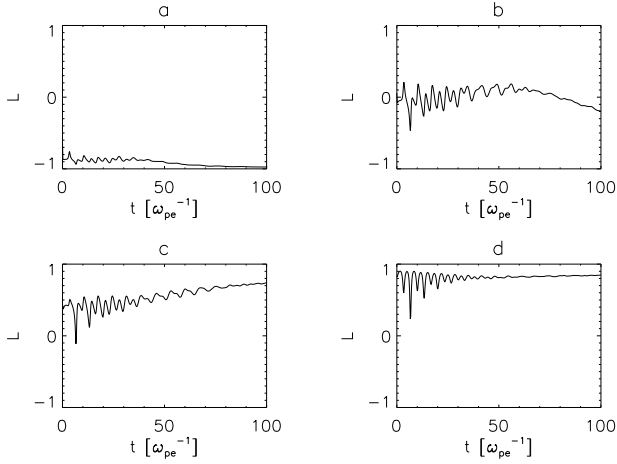}
\caption{\label{fig:linpol} Degree of linear polarization $L$ as a function of time for a) $\theta=15^{\circ}$ b) $\theta=45^{\circ}$ c) $\theta=60^{\circ}$ d) $\theta=90^{\circ}$ }
\end{figure}

The figure shows the time development of the ratio $w(t)/E^{beam}_{kin}(0)$ for different beam injection angles. We can see that the mechanism has a typical efficiency of $\approx 10^{-3}$-$10^{-2}$. Most notable is that, for our parameters, a smaller injection angle $\theta$ results in a higher efficiency. Note that this is not in contradiction with Fig.\ref{fig:IoverTheta}. Here, we take all electromagnetic field components into account, whereas, in Fig.\ref{fig:IoverTheta}, we focused solely on the peak intensity of the electric $y$-component, namely the wavelet transform of $E_y$, and showed a direct relation to the $y$-component of the kinetic energy of the beam. Fig.\ref{fig:IoverTheta} is also in line with the previously presented time-distance plots Figs.\ref{fig:tdds15}-\ref{fig:tdds90}. Comparing the scales in the respective color bars shows that the parallel component of the electric field, hence the electrostatic wake of the beam, is dominant for the presented cases (except of course for the $\theta=90^{\circ}$-case, where the parallel component vanishes).
Fig.\ref{fig:EmofEk} shows the pulsating nature of the early stages of the emission generation. The pulsation can be associated with the electromagnetic part of the emission, because it is generated in all cases. In the $\theta=90^{\circ}$-case (Fig.\ref{fig:EmofEk}d), the initially generated electromagnetic emission stabilizes at $\approx 10^{-3} E^{beam}_{kin}(0)$. In the other cases, we see that eventually the electrostatic wake of the beam overpowers the electromagnetic part significantly and subsequently causes an increase in the total field energy up to $\approx 10^{-2} E^{beam}_{kin}(0)$ in the $\theta=15^{\circ}$-case (Fig.\ref{fig:EmofEk}a).\\
The degree of linear polarization $L$ is defined similarly to Ref.\cite{2012ApJ...752...60Y}

\begin{eqnarray}
  L = \frac{w_{\perp} - w_{\parallel}}{w_{\parallel} + w_{\perp}} %
\end{eqnarray}

and its time development is presented in Fig.\ref{fig:linpol}.

\begin{figure}
\includegraphics[scale=0.5]{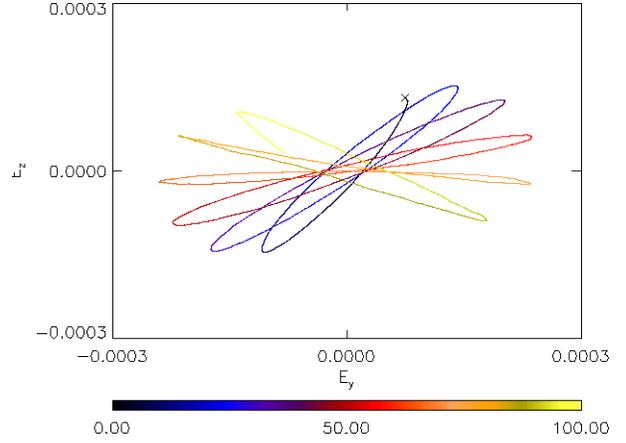}
\caption{\label{fig:yz-pol} Polarization in the perpendicular ($y$,$z$-)plane as a function of time for $\theta=90^{\circ}$. The $x$-axis points out of the paper plane. The black 'x' marks the starting point, color corresponds to time in units of $\omega_{pe}^{-1}$. Electric field components are given in units of $\omega_{pe}cm_e/e$. }
\end{figure}

Fig.\ref{fig:linpol} shows a strong dependence of $L$ with respect to the injection angle. As shown in Fig.\ref{fig:EmofEk}, the parallel component of the field energy becomes dominant for small injection angles, but is negligible in the $\theta=90^{\circ}$-case. Not surprisingly, we see a very strong degree of linear polarization for this case.\\
Concentrating on $\theta=90^{\circ}$, we plot the time evolution of the electric field in the perpendicular ($y$,$z$-)plane in Fig.\ref{fig:yz-pol}. In the figure, the $x$-axis points out of the plane. The time development is then shown by a change of pixel color, therefore we see that the wave is left-hand polarized.

A cyclotron maser in our parameter space ($\omega_{ce}/\omega_{pe} \ll 1$) is expected to generate waves in the z-mode, at frequencies that are harmonics of the cyclotron frequency \cite{1986ApJ...307..808W}. The polarization of the z-mode is left-handed for $\omega<\omega_{pe}$ and right-handed for $\omega>\omega_{pe}$ \cite{2000PhPl....7.3167W}. This is in line with our result, as the stable wave packet was generated below the plasma frequency, and is therefore expected to show left-hand polarization.

\section{\label{sec:conc}Conclusion}
In this paper, we study further the details of the non-gyrotropic beam driven emission mechanism, first outlined in Ref.\cite{2011PhPl...18e2903T}. We carried out 1.5D PIC simulations of a super-thermal electron beam being injected into a magnetised, maxwellian plasma. We investigated the role of the injection angle (pitch angle) of the beam with respect to the constant background magnetic field. \\
The evolution of spatial wavelet transforms of $E_y$ was demonstrated in movies. The movies show that, initially, pulsating emission well below the local plasma frequency is generated. Waves of such low frequencies cannot propagate in plasma, thus, in order for the waves to escape, a mode conversion has to take place. In the movies, mode conversion is identified by the formation of a stable wave packet, which drifts to higher frequencies, propagating only very slowly from the injection region. When the wave packet reaches frequencies of the order of the plasma frequency, propagation accelerates. The frequency drift, however, does not stop at this point, but only when the wave packet frequency reaches the frequency that corresponds to the plasma frequency at the beam injection point. Once this frequency is reached, the wave packet propagates without any frequency drift.\\
Four cases of different density gradients were studied: a) a constant background density b) a weak gradient c) the default case as used by Ref.\cite{2011PhPl...18e2903T} and d) a strong gradient. The characteristics of the emission mechanism were not changed by a different density profile. However, a steeper gradient means that the wave packet can reach frequencies that allow for propagation earlier, while the weak gradient has the opposite effect. In case of the constant background, no emission could escape the beam injection region, as the necessary frequencies were not reached within the simulated time. Thus we conclude, that the mode coupling, namely the cyclotron emission coupling to the allegedly z-mode, is facilitated by the \textit{density gradient}.\\
We found that the intensity of the wavelet transform of $E_y$ was proportional to the $y$-component of the kinetic energy of the injected beam. There was no other significant influence of the variation of the beam injection angle $\theta$. However, in the $\theta=90^{\circ}$-case, there was no electrostatic Langmuir wave signal found. This can be explained by the lack of a positive gradient in the distribution function with respect to parallel momentum. Nevertheless, generation of electromagnetic waves was evident. This sets the presented mechanism apart from conventional mechanisms, presented in the introduction. \\ 
Further, movies were generated to show the evolution of the distribution function in phase space. On one hand, when plotted in cartesian coordinates, gyration of the beam around the magnetic field could be seen; on the other hand, when plotted on a plane of parallel and perpendicular momentum axes, positive gradients as well as subtle dynamics of the distribution function was observed. The key finding is that, indeed, the requirement for the cyclotron maser instability, $\frac{\partial f}{\partial p_{\perp}}>0$, is fulfilled throughout the simulation. Cyclotron masers in overdense plasmas ($\omega_{ce}/\omega_{pe} \ll 1$) are expected to generate waves in the z-mode, at harmonics of the electron cyclotron frequency \cite{1986ApJ...307..808W}.\\
We would like to comment, why we think there is no continuous generation of cyclotron maser emission, despite the fact that $\frac{\partial f}{\partial p_{\perp}}>0$ throughout the simulation. The quasi-linear relaxation, i.e. the plateau formation shown in Ref.\cite{1999JGR...10410317P}, that shuts off the instability, happens on the time scale of the inverse growth rate. The growth rate calculation of our parameter space will be presented elsewhere. The reason why the emission generation is not continuous is, that we do not replenish the beam, as it is only injected at $t=0$. In solar flares, of course, the electron beam injection will have temporal (and spatial) extent.\\
The distribution function shows subtle dynamics throughout the simulation. Some correlations can be seen between the evolution of the wavelet transforms and the distribution function, such as in the initial pulsation phase as well as at the wave packet ejection point. This needs to be investigated further.\\
The degree of linear polarization of the escaping wave packet was strongly dependent on the injection angle. For the $\theta=90^{\circ}$-case, a left-handed polarization of the emitted wave was found. This is in line with z-mode wave generation, as the stable wave packet is actually formed below the plasma frequency and the polarization of the z-mode for $\omega<\omega_{pe}$ is left-handed \cite{2000PhPl....7.3167W}. At present we do not have definite proof that the generated emission is a z-mode. Such proof can be obtained by producing different numerical runs and producing $(\omega,k)$-pairs and compare these to predictions of dispersion relations as in Ref.\cite{2000PhPl....7.3167W} which presents a comprehensive study, but not exactly for the same parameters, that are used here. \\
The total generated electromagnetic field energy was of the order of $10^{-3}$ - $10^{-2}$ of the initial kinetic energy of the beam and varied as the pitch angle was altered.\\
The presented mechanism is relevant to solar type III radio bursts, as it is triggered by the \textit{perpendicular} component of the electron beam, while the parallel component merely generates Langmuir waves. The present study focuses on the injection of a single beam as an initial value problem (i.e. without replenishing the beam), but in flares many beams are generated and due to the outward propagation, the cyclotron maser is triggered along the beam trajectory, which mode converts on the density gradient into the z-mode, thus generating electromagnetic emission of decreasing frequencies, that give the characteristic shape of the dynamical spectrum for type III bursts. Ref.\cite{2009SoPh..259..255R} indicates that beam pitch angles are a few degrees. This means that Langmuir waves would be generated, and yet, as shown in Ref.\cite{PPL}, Langmuir waves play no role in EM emission generation, because, in 1.5D, plasma emission is not possible. Reality is - of course - not 1.5D and, in higher dimensions, plasma emission will switch on; as well as the antenna mechanism, if density cavities are present. Additional studies are needed to measure relative strength/importance of the different mechanisms in different physical situations.
Further analysis may also help to better understand the mechanism presented in this study. Such analysis may include the determination of the mode of the escaping emission as well as detailed analysis of the mode coupling process. \\
\\
The authors are financially supported by the HEFCE-funded South East Physics Network (SEPNET). D.T.'s research is supported by The Leverhulme Trust Research Project Grant RPG-311 and STFC Grants ST/J001546/1 and ST/H008799/1.\\
{\bf note added in proofs:} After this work was complete, we became aware of the following: (i) when a case of $T_{background}=3\times10^8$ is studied, eliminating the electron cyclotron maser (ECM) instability as $\frac{\partial f}{\partial p_{\perp}}<0$, $\forall {\bf p}$, EM emission identical to the one in this study still occurs; (ii) when a ring-shaped (in $p_y$ and $p_z$) beam distribution is considered, implying ${\bf j_{\perp}}=0$, no EM emission is generated. These findings indicate that both effects - the ECM instability and EM emission from transverse currents - are present (and competing). It is probable that the ECM instability growth rate is so small that it cannot develop by the end of the simulation. Calculation of the growth rates commensurate to physical parameters of type III bursts and the ring distribution will be published elsewhere.

%\bibliography{PoP}% Produces the bibliography via BibTeX.

\end{document}